\documentclass[%
nofootinbib,
 amsmath,amssymb,
 aps,
]{revtex4-2}

\usepackage{graphicx}
\usepackage{dcolumn}
\usepackage{bm}
\usepackage[hidelinks]{hyperref}
\usepackage{subcaption}

\begin{document}


\title{An Improvement of {\bf AmpRed}: Analytic Continuation of Complex Integrals}

\author{Wen Chen}
 \email{chenwenphy@gmail.com}
\affiliation{%
State Key Laboratory of Nuclear Physics and Technology, Institute of Quantum Matter, South China Normal University, Guangzhou 510006, China
}%
\affiliation{%
Guangdong Basic Research Center of Excellence for Structure and Fundamental Interactions of Matter, Guangdong Provincial Key Laboratory of Nuclear Science, Guangzhou 510006, China
}%


\date{\today}

\begin{abstract}
The {\bf AmpRed} package has been updated with an improved method for analytic continuation of complex integrals. Compared to the previous version, the new implementation significantly enhances computational efficiency for evaluating complex integrals.
\end{abstract}

\maketitle


{\bf AmpRed}~\cite{Chen:2024xwt} is a package designed for the semi-automatic calculations of multi-loop Feynman amplitudes\footnote{Available at: \url{https://gitlab.com/chenwenphy/ampred.git}}. It implements the methods developed by the author and collaborators in a series of papers~\cite{Chen:2019mqc,Chen:2019fzm,Chen:2020wsh,Chen:2023hmk}. Modern techniques for multi-loop calculations involve reducing Feynman integrals to master integrals~\cite{Tkachov:1981wb,Chetyrkin:1981qh,Laporta:2000dsw} and subsequently evaluating these master integrals. For details, we refer to ref.~\cite{Chen:2024xwt} and references therein. The main feature of {\bf AmpRed} is that it implements all these operations through the Feynman-parameter representation. In this paper, we focus on the calculations of master integrals.

A general parametric Feynman integral takes the form\footnote{Cut integrals~\cite{Chen:2020wsh} are not considered here.}:
\begin{equation}\label{eq:ParInt}
\begin{split}
I(\lambda_0,\lambda_1,\ldots,\lambda_n)=\frac{\Gamma(-\lambda_0)}{\prod_{k=1}^{n+1}\Gamma(\lambda_k+1)}\int \mathrm{d}\Pi^{(n+1)}\mathcal{F}^{\lambda_0}\prod_{k=1}^{n+1}x_k^{\lambda_k}\equiv\int \mathrm{d}\Pi^{(n+1)}\mathcal{I}^{(-n-1)}~.
\end{split}
\end{equation}
Here $\mathrm{d}\Pi^{(n+1)}\equiv\prod_{i=1}^{n+1}\mathrm{d}x_i\delta(1-\mathcal{E}^{(1)}(x))$ is the integration measure, with $\mathcal{E}^{(i)}(x)$ being a positive definite homogeneous function of $x$ of degree $i$. $\mathcal{F}$ is a homogeneous polynomial of $x$. For loop integrals, $\mathcal{F}$ is related to the well-known Symanzik polynomials $U$ and $F$ through $\mathcal{F}=F+Ux_{n+1}$. Due to the $i0^+$ prescription of Feynman propagators, $\mathcal{F}$ has a negative imaginary part.

{\bf AmpRed} employs the iterative algorithm developed in ref.~\cite{Chen:2023hmk} to calculate master integrals. Specifically, we introduce an auxiliary scale $y$ into the integral $I$ in eq.~(\ref{eq:ParInt}) by inserting a trivial integral $\int\mathrm{d}y~\delta(y-\mathcal{E}^{(0)})$. That is,
\begin{equation}
    I=\int \mathrm{d}\Pi^{(n+1)}\mathrm{d}y~\delta(y-\mathcal{E}^{(0)}(x))\mathcal{I}^{(-n-1)}~.
\end{equation}
In practice, we choose
\begin{equation}
    \mathcal{E}^{(0)}=\frac{x_i}{x_j}~.
\end{equation}
Then we get
\begin{equation}
\begin{split}
    I=\int\mathrm{d}y\int\mathrm{d}\Pi^{(n)}~x_j\left.\mathcal{I}^{(-n-1)}\right|_{x_i=yx_j}=\frac{\Gamma(\lambda_i+\lambda_j+2)}{\Gamma(\lambda_i+1)\Gamma(\lambda_j+1)}\int\mathrm{d}y~y^{\lambda_i}I_y~,
\end{split}
\end{equation}
where
\begin{equation}
    I_y=\frac{\Gamma(-\lambda_0)}{\prod_{k\neq i}\Gamma(\lambda_k^\prime+1)}\int \mathrm{d}\Pi^{(n)}\mathcal{F}_y^{\lambda_0}\prod_{k\neq i}x_k^{\lambda_k^\prime}~,
\end{equation}
with $\mathcal{F}_y\equiv\left.\mathcal{F}\right|_{x_i=yx_j}$, $\lambda_j^\prime=\lambda_i+\lambda_j+1$, and $\lambda_k^\prime=\lambda_k$ for $k\neq j$. The obtained integral $I_y$ is further calculated by using the differential-equation method~\cite{Kotikov:1990kg,Remiddi:1997ny,Gehrmann:1999as}.

For complex integrals, a problem with this method is that the integral $I_y$ may have some branch points in $y$, and it is generally not easy to determine the physical branch. While a solution was provided in ref.~\cite{Chen:2023hmk} (see sec.~3.4 therein), it may greatly reduce efficiency.

{\it An observation is that the branch is fixed by the $-i0^+$ prescription of the $\mathcal{F}$ polynomial if the coefficients of the monomials $y^a$ in $\mathcal{F}_y$ are positive (or negative) definite. Thus, if we choose the pair $\{x_i,~x_j\}$ in $\mathcal{E}^{(0)}$ such that the coefficients of the monomials $x_i^a$ in $\mathcal{F}$ are positive (or negative) definite, the branch of $I_y$ in $y$ is fixed. Notice that such a choice is always possible for loop integrals, because the coefficient of $x_{n+1}$, which is just the first Symanzik polynomial $U$, is always positive definite. Hence, we can calculate complex integrals with the same method as real integrals, and the efficiency is thus close to that of real integrals.}

\begin{figure}[]
 \centering
 \includegraphics[width=0.9\linewidth]{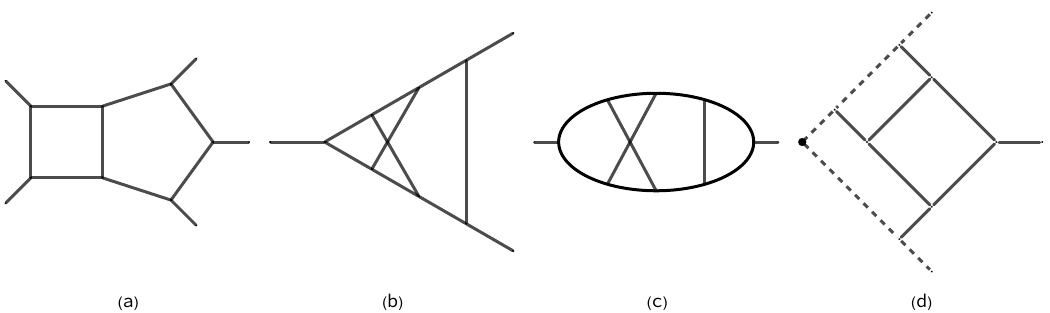}
 \caption{\label{fig:diagr}Diagrams corresponding to the test integrals. All the internal lines are massless, and dashed lines represent Wilson lines.}
\end{figure}

We benchmark the method on some integrals up to four loops (including both real and complex integrals). The corresponding diagrams are shown in Fig.~\ref{fig:diagr}. The first three integrals are normal loop integrals, and the last integral is an integral with two Wilson lines (linear propagators). The detailed calculations can be found in the example notebook {\tt "AmpRed/examples/MasterIntegrals.nb"}. As a comparison, we also calculated these integrals by using {\tt AMFlow}~\cite{Liu:2022chg}. Both calculations used {\tt KIRA}~\cite{Maierhofer:2017gsa,Kauers:2008zz,Klappert:2020nbg} to solve the integration-by-parts (IBP) identities. Computation times of these calculations are shown in Tab.~\ref{tab:timing}. As can be seen from the table, while {\bf AmpRed} is less efficient for the two-loop integral in Fig.~\ref{fig:diagr}a, it is much more efficient for the rest three- and four-loop integrals. And {\tt AMFlow} failed to calculate the integral with Wilson lines due to the lack of memory~(As a comparison, the memory used by {\bf AmpRed} is about 200GB).

\begin{table}[]
    \centering
    \begin{tabular}{ccccc}
    \multicolumn{1}{l}{CPU:~} & \multicolumn{4}{l}{AMD EPYC 7R32 / 2.8Ghz-3.3Ghz}\\
    \multicolumn{1}{l}{RAM:~} & \multicolumn{4}{l}{520 GB}\\
    \multicolumn{1}{l}{Threads:~}& \multicolumn{4}{l}{24}\\
    \multicolumn{1}{l}{IBP solver:~}& \multicolumn{4}{l}{\tt KIRA2.3}\\
    \multicolumn{1}{l}{Precision goal:~}& \multicolumn{4}{l}{20}\\
    ~\\
    \hline
    diagram & a & b & c & d\\
    \hline
    {\bf AmpRed} & ~$5.27\times10^3$~ & ~$2.86\times10^3$~ & ~$3.39\times10^4$~ & ~$1.54\times10^4$~\\
    \hline
    {\tt AMFlow} & ~$2.70\times10^3$~ & ~$8.64\times10^3$~ & ~$1.31\times10^5$~ & ~?~\\
    \hline
    \end{tabular}
    \caption{Computation times (in seconds) for the test integrals in Fig.~\ref{fig:diagr}.}
    \label{tab:timing}
\end{table}

Notice that we calculate all these integrals in the Euclidean region, because the second Symanzik polynomials $F$ of these integrals have more positive terms in the Euclidean region. This approach is valid because $F$ is homogeneous in the Lorentz invariants (including all the Mandelstam variables and the squared masses). We collectively denote all the Lorentz invariants by $s$. A parametric integral $I$ is a function of $s$. For real $s$, it is easy to get
\begin{equation}
    I(-s)=e^{-i\pi\left(\lambda_0+\lambda_{n+1}+1\right)}I(s)^*~.
\end{equation}

For multi-scale integrals (like the one in Fig.~\ref{fig:diagr}a), instead of calculating them by brute force, {\it we recommend computing them using the differential-equation method with the boundary chosen at a singular point} (See the example notebook {\tt "AmpRed/examples/Banana.nb"}).

In summary, we update the package {\bf AmpRed} by improving the method of analytic continuation. Consequently, the efficiency in calculating complex integrals is significantly improved. Benchmarks on calculating several non-trivial integrals up to four loops are provided.

\acknowledgments

This work is supported by National Natural Science Foundation of China~(NNSFC) under Grant No. 12405095 and Guangdong Major Project of Basic and Applied Basic Research~(No. 2020B0301030008).

\bibliography{refs}

\end{document}